\documentclass{article}
\usepackage{sw20spie}
\usepackage{graphicx}
\usepackage{amsmath,amssymb,bbm}

\begin{document}

\title{\textbf{Comment on "Three-dimensional hydrodynamic simulations of the
combustion of a neutron star into a quark star"}}
\author{  {\Large M. I. Krivoruchenko and B. V. Martemyanov} \vspace{15pt} \\
Institute for Theoretical and Experimental Physics$\mathrm{,}$ B.
Cheremushkinskaya 25\\
117218 Moscow, Russia }
\date{}
\maketitle

\begin{abstract}
If strange matter is absolutely stable, the ordinary nuclei decay to
strangelets, while neutron stars convert into strange stars. Lifetimes of
the ordinary nuclei are constrained experimentally to be above $\sim 10^{33}$
years, while lifetimes of the metastable neutron stars depend on the neutron
star masses and can exceed the age of the Universe. As a consequence, the
neutron stars and the strange stars can coexist in the Universe. We point
out that numerical simulations of the conversion of neutron stars to strange
stars, performed by M. Herzog and F. K. R\"opke in Phys. Rev. D \textbf{84},
083002 (2011), are focused on a region in the parameter space of strange
matter, in which low-mass neutron stars and strange stars are coexistent,
whereas massive neutron stars are highly unstable and short lived on an
astronomical timescale.
\end{abstract}

\vspace{50pt}


\pagebreak \setcounter{page}{1} 

In 1960 Ambartsumyan and Saakyan \cite{AMBA60} pointed out that production
of hyperons becomes energetically favorable inside neutron stars and
calculated the composition and the equation of state (EoS) of cold baryonic
matter. The existence of quark stars was conjectured by Ivanenko and
Kurdgelaidze \cite{Ivan65}. Quark stars whose interior consists of strange
quark matter (SQM) were discussed by Itoh in 1970 \cite{ITOH70}.
The next year Bodmer \cite{BODM71} discussed the possible existence
of absolutely stable SQM, the binding energy of which at zero external pressure
could be greater than that of ordinary nuclei.
This idea has attracted interest only much later,
after the publication of Witten's (1984) seminal paper~\cite{WITT84}.
Detailed calculations in the framework of
the MIT bag model, made by Farhi and Jaffe \cite{FARH84}, confirmed that for
an admissible range of parameters the SQM is bound.

Existence of the absolutely stable SQM does not contradict to the fact that
ordinary nuclei are composed of nucleons. The critical density of the phase
transitions with conservation and non-conservation of strangeness, i.e.,
with turned off and turned on weak interaction, differ only by a numerical
factor, so the bound SQM automatically implies that critical density of the
phase transition with the conservation of strangeness is relatively low.

At the saturation density, the nuclear matter is still in the baryonic
phase. From this condition, Farhi and Jaffe \cite{FARH84} derived a
restriction on the binding energy of the SQM and the vacuum pressure $B$.
For the quark-gluon coupling constant $\alpha_c = 0 $ and the strange quark
mass $m_s=100$ MeV they found $E/A > m_N - 90$ MeV, where $m_N$ is nucleon
mass, and
\begin{equation}
B^{1/4} > 145 \; \text{MeV.} \;\;\;\; \mathtt{(stability~of~nuclei)}
\;\;\;\;\;\;\;\;\;\;\;\;
\end{equation}

Stronger constraints follow from the existence of neutron stars. If SQM is
absolutely stable, the phase transition to quark matter with the
conservation of strangeness leads, on a weak interaction time scale, to the
conversion of quark matter to SQM and the eventual slow (deflagration) or
fast conversion of a neutron star into a strange star. Such a process is
followed by neutrino burst, which can be detected by terrestrial
observatories. Neutrino bursts accompanying the slow combustion of neutron stars into quark stars
can have duration up to 5 min. with average neutrino energy of a few
MeV. \cite{Mart94} 1-minute intervals on the historical data were analyzed at
the Baksan scintillation telescope~\cite{BST}; no statistically significant
excess of neutrino events was found.
10-second signals are expected from supernova explosions and also from
the conversion of neutron stars into quark stars
in the scenarios in which the phase transition leads to gravitational instability. \cite{MIKR87}
Given that neutrino signal is not confirmed by an observation of supernova
in optics, x-rays or gamma rays, in-depth analysis of neutrino spectrum and the time evolution of
the signal is required to discriminate between scenarios.
In this respect, the long-duration neutrino signals provide a cleaner signature for the identification of "quark-novae".
10-second intervals were analyzed at the Sudbury Neutrino Observatory
\cite{SNO}, also with zero result.

If the conjecture on the absolutely stable SQM is correct, then all compact
stars are either neutron stars or strange stars. The neutron stars are
metastable with respect to conversion to strange stars, and their central
density is lower than the critical density of the phase transition with the
conservation of strangeness.

In observational astrophysics, there are ample indications that compact
objects are neutron stars rather than strange stars. The existence of
neutron stars provides stronger constraints on the parameters of the MIT bag
model, compatible with the absolutely stable SQM. These constraints can be
further improved by considering conditions inside newly born hot
protoneutron stars during the first seconds after the core collapse
supernovae. Such limits are derived in Refs. \cite{MIKR87,MIKR91a,MIKR91b}
For $\alpha_c = 0$, e.g., the absence of phase transition with the
conservation of strangeness inside a protoneutron star with the baryon rest
mass of $1.4~M_{\odot}$ implies
\begin{equation}
B^{1/4} > 155 \; \text{MeV.} \;\;\;\; \mathtt{(stability~of~neutron~stars)}
\end{equation}

In recent years, pulsars with masses $\sim 2.0~M_{\odot}$ have been
reported. \cite{Demo10,Anto12} Such pulsars, if they are neutron stars, have
a high central density and more favorable conditions for conversion into
strange stars. A neutron star with mass $\sim 2.0~M_{\odot}$ rules out three
softest EoS of nuclear matter out of the six ones examined in Refs.~\cite%
{MIKR91a,MIKR91b} The stiff EoS \cite{BARO85,PAND75a,PAND75b} still do not
contradict the absolutely stable SQM, although its binding energy must be
small. For $\alpha_c = 0$ and $m_s=100$ MeV, the binding energy of less than
30 MeV is allowed in the model of Baron, Cooperstein and Kahana \cite{BARO85}%
, while in the models of Pandharipande and Smith \cite{PAND75a,PAND75b} the
SQM is, for this particular choice of parameters, unbound.

An evidence that SQM is unbound comes from the relatively high crossover
temperature of QCD. The absolutely stable SQM requires for two-flavor quark
mater $T_{c}(n_f = 2) < 122 \pm 7$ MeV \cite{Kond91}, whereas numerous
lattice data give $T_{c}(n_f = 2) = 175 \pm 10$ MeV. \cite{Satz11}

It is presently unknown whether strange stars exist at all. On the other
hand, it has been discussed whether there do exist undoubted observable
indications that at least some of the compact stars are not strange stars
\cite{CALD91,Balb92,JMad04}. The constraints of Refs. \cite%
{MIKR87,MIKR91a,MIKR91b} are valid provided one can find in the Universe at
least one neutron star with a mass above $1.4~M_{\odot}$.

In Ref.~\cite{HERZ11} results of the numerical simulation of combustion of a
neutron star into a strange star are reported. The conversion to the
absolutely stable SQM is found to be turbulent for a substantial part of the
parameter space of the MIT bag model. This interesting analysis oversteps
the line of the coexistence of massive neutron stars and strange stars:

Given a neutron star of mass $1.4~M_{\odot}$, the horizontal lines in
Figs.~1 and 2 of Ref.~\cite{HERZ11} should be moved up to the level of 155
MeV. The left two-thirds of Fig.~3, four out of the five rows of Table 3,
three of the four curves in Fig.~4, all the Fig.~5, four of the six columns
of Table~2, four of the five curves in Fig.~6, all the Fig.~7, and Sections
5B and 5C are also excluded.

Almost all scenarios described in Ref.~\cite{HERZ11} belong to a universe
without the massive neutron stars.

During the time that has elapsed since the publications of Bodmer~\cite%
{BODM71} and Witten~\cite{WITT84}, there arose a disagreement with lattice
models concerning crossover temperature of QCD, while experimental searches
for stable strange matter in the laboratory and astrophysics did not yield
positive results. Nevertheless, strange quark matter still deserves careful
study, for it has important physical implications. A strong argument in
favor of the existence of strange quark matter would be the observation of a
long-duration soft neutrino burst accompanying combustion of a neutron star
into a quark star. Observation of two pulsars with equal masses and
different radii would provide an indirect evidence for coexistence of
neutron stars and strange stars. Another point worth noting is that since
low-mass strange stars are bound by the strong force, their rotation speed
is not constrained by the Kepler frequency. A period of rotation $\lesssim
0.5$ milliseconds would indicate a low-mass strange star. Eventually, one
can hope that fast progress in lattice gauge theories on their part will
help to also describe quantitatively cold baryonic matter.

The authors gratefully acknowledge correspondence from N. Itoh.

\newpage

\renewcommand{\thesection}{}

\end{document}